\documentclass[conference]{IEEEtran}
\usepackage{cite}
\ifCLASSINFOpdf
\else
\fi

\usepackage{tabularx}

\usepackage{url}
\usepackage[hidelinks]{hyperref}
\usepackage{svg}
\usepackage{soul}

\begin{document}
\title{Designing Virtual Reality Games for Grief: A Workshop Approach with Mental Health Professionals}


\author{\IEEEauthorblockN{Amina Kobenova\IEEEauthorrefmark{1}, Piper Stickler\IEEEauthorrefmark{1}, Thais Alvarenga\IEEEauthorrefmark{2}, Sri Kurniawan\IEEEauthorrefmark{1}}
\IEEEauthorblockA{\IEEEauthorrefmark{1}Department of Computational Media,
University of California, Santa Cruz\\
Santa Cruz, CA, USA\\
\IEEEauthorrefmark{2}School of Literature, Media, and Communication,
Georgia Institute of Technology\\
Atlanta, GA, USA\\
\IEEEauthorrefmark{1}akobenov@ucsc.edu,
\IEEEauthorrefmark{1}pstickle@ucsc.edu,
\IEEEauthorrefmark{2}thais@gatech.edu,
\IEEEauthorrefmark{1}skurnia@ucsc.edu}}

% \author{
% % First row of authors
% \IEEEauthorblockN{Amina Kobenova}
% \IEEEauthorblockA{
% \textit{Department of Computational Media} \\
% \textit{University of California, Santa Cruz}\\
% Santa Cruz, CA, USA \\
% akobenov@ucsc.edu}
% \and
% \IEEEauthorblockN{Piper Stickler}
% \IEEEauthorblockA{
% \textit{Department of Computational Media} \\
% \textit{University of California, Santa Cruz}\\
% Santa Cruz, CA, USA \\
% pstickle@ucsc.edu}
% \and 
% % Second row of authors
% \IEEEauthorblockN{Thais Alvarenga}
% \IEEEauthorblockA{
% \textit{School of Literature, Media, and Communication} \\
% \textit{Georgia Institute of Technology}\\
% Atlanta, GA, USA \\
% thais@gatech.edu}
% \and
% \IEEEauthorblockN{Sri Kurniawan}
% \IEEEauthorblockA{
% \textit{Department of Computational Media} \\
% \textit{University of California, Santa Cruz}\\
% Santa Cruz, CA, USA \\
% skurnia@ucsc.edu}
% }

\maketitle

\begin{abstract}

Although serious games have been increasingly used for mental health applications, few explicitly address coping with grief as a core mechanic and narrative experience for patients. Existing grief-related digital games often focus on clinical training for medical professionals rather than immersive storytelling and agency in emotional processing for the patient. In response, we designed \textit{Road to Acceptance}, a VR game that presents grief through first-person narrative and gameplay. As the next phase of evaluation, we propose a workshop-based study with 12 licensed mental health professionals to assess the therapeutic impacts of the game and the alignment with best practices in grief education and interventions. This will inform iterative game design and patient evaluation methods, ensuring that the experience is clinically appropriate. Potential findings can contribute to the design principles of grief-related virtual reality experiences, bridging the gap between interactive media, mental health interventions, and immersive storytelling.

\end{abstract}


% \fi
\IEEEpeerreviewmaketitle

\section{Introduction}

Grief is a complex, personal, yet ubiquitous human experience. Mental health and psychology scholars describe grieving as a complex and multi-sequential process that can span from several months to years \cite{gustafson1989grief, kubler2009five, averill1968grief}. Psychological research on grief has proposed several theoretical models to understand the emotional and cognitive processes involved in loss, such as Kübler-Ross’s five-stage model (denial, anger, bargaining, depression, acceptance) \cite{kubler2009five}. However, contemporary perspectives suggest that grief is non-linear and highly individualized. The dual-process model of coping with bereavement \cite{schut1999dual} further expands on this by distinguishing between loss-oriented and restoration-oriented coping strategies.

Grief interventions and coping techniques are widely explored in psychotherapy studies \cite{johannsen2019psychological, mason2020complicated, neimeyer2015techniques}, with increasing explorations of grief in human-computer interaction (HCI) and VR interventions \cite{botella2008treatment, quero2019adaptive, marchioro2024processing}. Despite its ubiquity, grief-related interventions are rarely explored and evaluated through the lens of gamified narratives.

% Despite its ubiquity, grief and related interventions are rarely explored through the lens of gamified narratives. Existing interventions often propose using novel technologies, such as generative artificial intelligence (GenAI) and VR, to recreate memories and experiences \cite{chen2024using, pizzoli2023virtual} due to the design affordances and psychological effects of being immersed in 3D virtual spaces \cite{han2024virtual, zhang2024designing, marchioro2024processing}. However, little existing work explores embodied VR experiences for grief interventions and multi-sequential coping through serious games.

To address this gap, we propose a study that leverages an existing VR game, \textit{Road to Acceptance}, to gain insight from grief professionals, licensed mental health counselors and professionals, on gamified VR interventions. \textit{Road to Acceptance} was previously introduced as an extended abstract that detailed its narrative and mechanics \cite{alvarenga2024road}. This study expands previous work by focusing on its evaluation through expert feedback to answer the research questions:
\begin{itemize}
    \item \textbf{RQ1:} How do counselors perceive the VR game as a grief intervention tool?
    \item \textbf{RQ2:} Does the game align with best practices in grief counseling?
    % \item \textbf{RQ3:} How do counselors envision using novel technology tools, such as VR games and GenAI, in their practice?
\end{itemize}

Our work contributes to the broader field of serious games and mental health. First, the study explores how VR can be used beyond traditional therapy to support grief awareness. Second, VR experiences like \textit{Road to Acceptance} could be incorporated into grief counseling programs and mental health education, making grief support more accessible. Lastly, the study may highlight the potential for interactive storytelling and gamified experiences to help individuals in non-clinical settings.
\section{Study Design Proposal}

While the gameplay mechanics of \textit{Road to Acceptance} have been described in detail in a prior extended abstract \cite{alvarenga2024road}, this paper focuses on proposing the study design and research methodology. The gameplay follows an episodic structure where players navigate key grief-related emotions defined by the five-stage model of Kübler-Ross \cite{kubler2009five} through interactive storytelling. Each stage (denial, anger, bargaining, depression, acceptance) presents micro-games that reinforce the emotional arc of grieving. Unlike traditional narrative-driven serious games, \textit{Road to Acceptance} leverages embodied VR interactions—such as symbolic actions (physical movement, releasing objects, negotiating with a non-playable character, revisiting memories)—to engage users in reflective and therapeutic activities.

The game underwent significant technical development and followed a playtesting methodology \cite{turkki2024indie} for the pilot assessment. The pilot playtests were conducted with 2 mental health specialists from a nonprofit community organization recruited from targeted outreach. The playtesters commented on the therapeutic potential of the game and proposed ideas for working with children and families to educate about grief.

We propose the study to evaluate \textit{Road to Acceptance} \cite{alvarenga2024road} with twelve (N=12) mental health and psychosocial professionals. While current studies on serious games and grief focus on educational and clinical outreach, we tailor our assessment to practitioners who will evaluate the game targeted to patients and the general public. The evaluation will follow a workshop approach, in which participants will share grief intervention techniques they follow in their practice, participate in the game experience, and speculate on design enhancements for similar games afterward.

Licensed counselors and mental health professionals will be recruited via professional networks, online outreach, and grief therapy associations. To gather the depth of participant responses and detailed feedback about the game experience, we intend to employ qualitative interviews \cite{cote2015depth} and a Likert-style questionnaire at the end of the experience to measure the overall usability of our system \cite{olsen2011serious}. The recruitment will follow immediately after the university's Institutional Review Board (IRB) approval. Each study is expected to last 60-90 minutes and structured as: (1) Participants receive a brief explanation of the study and the VR experience; (2) Participants engage with \textit{Road to Acceptance} using VR equipment; (3) Semi-structured interviews will explore their perceptions of the game, therapeutic potential, and ethical considerations; (4) Participants complete a short questionnaire using a Likert scale. The participants' interviews will undergo thematic analysis \cite{braun2024thematic, clarke2017thematic} completed by the research team. The findings of our proposed study will be discussed as a set of design implications and a framework recommendation for serious games exploring grief and coping with loss.

Our study proposal is aimed at practitioners working with grief in their intervention practices. Little is still known about how well serious games reflect grief processing with real patients and whether games avoid misinterpretation and retraumatization. As a result, it is important to consider how real patients might react to gamified interventions. Having mental health professionals evaluate the game minimizes the risk of any emotional distress caused by the experience.

Mental health research presents agency as an empowering framework to process trauma and loss \cite{lorimer2022improving, warner2020transforming}. Serious games in VR aim to design for user agency, interactivity, and feedback \cite{nguyen2025games, tanenbaum2010agency}. As such, gaming contextualizes agency to those coping with grief and learning about it. \textit{Road to Acceptance} and our research seek to broaden this avenue of work and contribute to ongoing efforts to design better therapeutic tools and technologies. 
\section{Future Work and Conclusion}

This study proposal examines a serious VR game for grief counseling in mental health interventions. We propose an evaluation with licensed professionals to assess playful approaches to traditional therapy. This aims to provide insights into the possibilities of VR to support awareness and emotional resilience regarding this universal human experience. 

In addition to serious games in VR, our approach can explore the potential of other emerging technologies, such as game artificial intelligence (AI), in mental health interventions. This can include AI-generated narratives or conversational agents as assistive features in technologically mediated therapeutic applications. Based on our proposed studies, future iterations may explore AI-assisted storytelling and agents to personalize therapy experiences in serious games.

This work seeks to encourage critical feedback from licensed professionals when designing emerging interventions around grief. Such evaluations ensure that these tools not only engage users but also responsibly support emotional awareness and the development of healthy coping skills through safe, thoughtfully crafted experiences.

\section*{Acknowledgments}

We would like to thank various collaborators and sponsors who have contributed to the development of this game and the study: CITRIS and the Banatao Institute at the University of California, Michael Matkin, Saara Korpela, Michael Allison, Ons Taktak, and our study participants.

\bibliographystyle{IEEEtran}
\bibliography{references}

% \input{appendix}

\end{document}